# Study of Ge Doped Garnet Type $Li_7La_3Zr_2O_{12}$ as Solid Electrolyte for Li-ion Battery Application

Muktai Aote[1], A.V. Deshpande [2]

[1,2] Department of Physics, Visvesvaraya National Institute of Technology, South Ambazari Road Nagpur,

Maharashtra 440010, India

[2] Corresponding Author E-mail Address [avdeshpande@phy.vnit.ac.in]

**Abstract**

$Li_{7-4x}Ge_xLa_3Zr_2O_{12}$ has been synthesized using the conventional solid-state reaction method by substituting Germanium (Ge) at the Li site, which increases the Li-ion vacancies and leads to an increase in conductivity with x varying from 0.05-0.20. The formation of cubic phase is confirmed by using XRD analysis. The surface morphology and elemental distribution have been studied using SEM characterization which gives the average particle size of the sample. The densities of the samples were calculated. For the confirmation of functional groups present within the sample, IR spectroscopy has been studied. The modulus and ac conductivity studies have also been studied. A complex impedance study has been done in the frequency range 20Hz -20 MHz. Increase in ionic conductivity by one order has been observed in the sample with x=0.10. The minimum value of 0.56 eV activation energy is associated with the highest conductivity value of 7.23 x $10^{-6}$ S/cm at room temperature. Thus increment in ionic conductivity at room temperature makes this material a promising solid electrolyte for future sustainable energy storage devices.

## 1. Introduction

As time passes, the world is encountering an energy crisis due to the high burn rate of fossil fuels. These kinds of energy sources are diminishing, building pressure on society for the upcoming generation's energy needs. The world is enriched with renewable energy sources; thus, many countries are trying to set up renewable energy plants, which contributes to the reduction in the usage of carbon-emitting fuels and promoting the era of electric vehicles. But the initiative that has been taken needs proper devices to store and convert energy. Some known energy devices are capacitors, fuel cells, batteries and solar cells. Ionic conduction is the main key factor of these devices[1,2]. Out of all these electrochemical devices, batteries are superior in energy density and duty cycle with the low self-discharge ability and lighter weight. Batteries are classified into two types, i.e. primary batteries, which are non-rechargeable and secondary batteries which are reusable after charging. The volumetric and gravimetric energy densities of secondary batteries are very high [3]. Thus it makes a secondary battery promising energy storage device [1]'[4]–[6].

As we are discussing batteries, the main components of the batteries are their anode, cathode, and electrolyte, which is the path for the flow of charge carriers from anode to cathode and vice versa. Mostly liquid electrolytes have been used in the batteries due to their high conductivity. But along with that, many issues have been aroused related to liquid electrolytes, such as leakage, which can lead to flammability and dendrite growth which hampers the conduction of ions [7], [8]. Thus to overcome these situations, the focus has been shifted towards developing solid electrolyte favoring ionic conduction. Many solid electrolytes have been studied, such as NASICON, LISICON, LIPON, Beta Alumina and $Li_3N$ type [9]–[12]. But all these have chemical instability, with lithium metal electrodes having low electrochemical potential windows. Also, they are difficult to prepare in pure form, affecting their ionic conductivity

Hence, a solid electrolyte is required to fit in all the criteria and overcome these previously mentioned problems. Solid electrolytes are made with non-flammable and volatile components, which helps in transport of lithium ions and prevent electronic transmission, simultaneously eliminating safety risk [13]–[15]. Here $Li_7La_3Zr_2O_{12}$, commonly known as LLZO, the garnet-type solid electrolyte, comes into the picture. It has very good thermal and chemical stability, with lithium electrodes having comparatively high conductivity and a high potential window [16]. LLZO possesses two structures which are tetragonal and cubic. But for the conduction of Li ions, the cubic phase is essential. The study on enhancing the ionic conductivity of LLZO at room temperature has been reported earlier by adding various dopants such as Al, Ba, Fe, Ga, Ta, and Bi in place of Li and Zr [17]–[22]. These dopants help in improving ionic conductivity by increasing Li-ion vacancies. Previously the effect of the partial substitution of Ge for Zr has been studied [23]. Thus in this work, the effect of substitution of supervalent Germanium (Ge) at the Li site has been studied to enhance ionic conductivity at comparatively lower sintering temperature.

## 2. Experimental Section

### 2.1 Sample Preparation

The series $Li_{7-4x}Ge_xLa_3Zr_2O_{12}$ (x= 0.05,0.10-0.20) with four compositions have been prepared using $Li_2CO_3$ (Merck, >99.9%), $La_2O_3$ (Sigma Aldrich, >99.99%), $ZrO_2$ (Sigma Aldrich, >99.9%) and $GeO_2$ (Sigma Aldrich, >99.99%) as preparative chemicals by conventional solid-state reaction method in an air atmosphere. The stoichiometric amounts of all the above chemicals have been taken in agate mortar and hand mixed thoroughly, followed by dry mixing for 2 h and wet mixing for 1 h using acetone as the dispersive agent. To compensate the loss of lithium during heating process, 10% excess lithium in the form of $Li_2CO_3$ has been added. The formed powder is then transferred into an alumina crucible and kept in a muffle furnace for calcination at $900^0C$ for 8 h. After the calcination, the powder was cooled at room temperature and again crushed into fine powder. To form the pellets of around 10 mm diameter and 1.5 mm thickness, the powder was uniaxially pressed under the pressure of 4 tons per $cm^2$ using a hydraulic press. Formed pellets were then kept in a bed of mother powder for sintering at $1050^0C$ for 7.30 h in a muffle furnace.

## 2.2 Characterizations

For the phase identification, the pellet sintered at $1050^0C$ was crushed and then examined by X-ray diffraction with RIGAKU diffractometer using Cu-kα having wavelength of 1.54 $A^0$ as a radiation source. The data were collected in the range of $10^0 - 80^0$ with a step size of 0.02 degrees and $2^0$/min scan speed. The densities of samples were determined by Archimedes principle with toluene as an immersion medium using K-15 Classic (K-Roy) instrument. Scanning electron microscopy (JSM-7600 F/JEOL) was done for the microstructural analysis and compositional study using an accelerating voltage of 15kV. The conductivity study and related modulus study have been carried out using a Novocontrol impedance analyzer in the frequency range 20Hz- 20KHz and temperature range varying from room temperature to $150^0C$. The silver paste was coated on both the faces of sintered pellets with a diameter of around 10mm and 1.3 mm thickness to maintain ohmic contact with the silver electrodes, which act as blocking electrodes for AC and DC conductivity measurements. DC polarization technique was carried out to calculate the ionic transport number using KEITHLEY 6512 programmable electrometer. The constant voltage of 1 V was applied to the silver electrodes in which the pellet was placed, and the corresponding current through the sample was measured in the nanometer range with time.

## 3. Results and Discussion

### 3.1 X-ray diffraction

Fig.1(a) shows the powder X-ray diffraction pattern of the $Li_{7-4x}Ge_xLa_3Zr_2O_{12}$ samples (x= 0.05,0.10-0.20), which are sintered at $1050^0C$ for 7.30 hours in a muffle furnace. All the recorded peaks agree with the peaks from JCPDS file no. 45.0109. The peaks are marked with the corresponding (hkl) planes. It confirms that all the samples possess cubic phase with space Group: Ia-3d, which is the primary requirement for developing highly conducting garnet type $Li_7La_3Zr_2O_{12}$. Among all the samples, the sample with x= 0.10 Ge shows accurate matching of more crystalline peaks with high intensity. This result is in agreement with the work by Huang [24]. It has been reported that, concentration of 1 wt% of Ge, which is nearly equal to 0.12 mol of Ge, helps to form and stabilize the cubic phase. This argument is well supported by the observed splitting of (321) and (420) peaks for samples with x= 0.05, 0.15, and 0.20 Ge. The splitting of these peaks may be attributed to the interference of the tetragonal phase (Space group: $I4_1/acd$) within the material. Along with the required phase, the $La_2Zr_2O_7$ phase has also been observed around $27^0$ (JCPDS file no. 17-0450). $La_2Zr_2O_7$ pyrochlore commonly occurs as an impurity phase in the synthesis of $Li_7La_3Zr_2O_{12}$. The occurrence of pyrochlore of lanthanum is due to the volatilization of lithium from the sample, which occurs during the sintering process [25]. Fig. 1(b) shows the shifting of (211) peak towards higher values of 2Θ. This result may be due to the increment of Ge content in the sample. The substitution of Ge in place of Li may cause a decrease in the lattice constant because the ionic radius of $Ge^{4+}$ is smaller than the ionic radius of $Li^+$, which results in a shrinking effect [26]. The average crystallite size for x= 0.10 Ge was calculated using the Debye Scherrer formula, and it was found to be 34.59 nm.

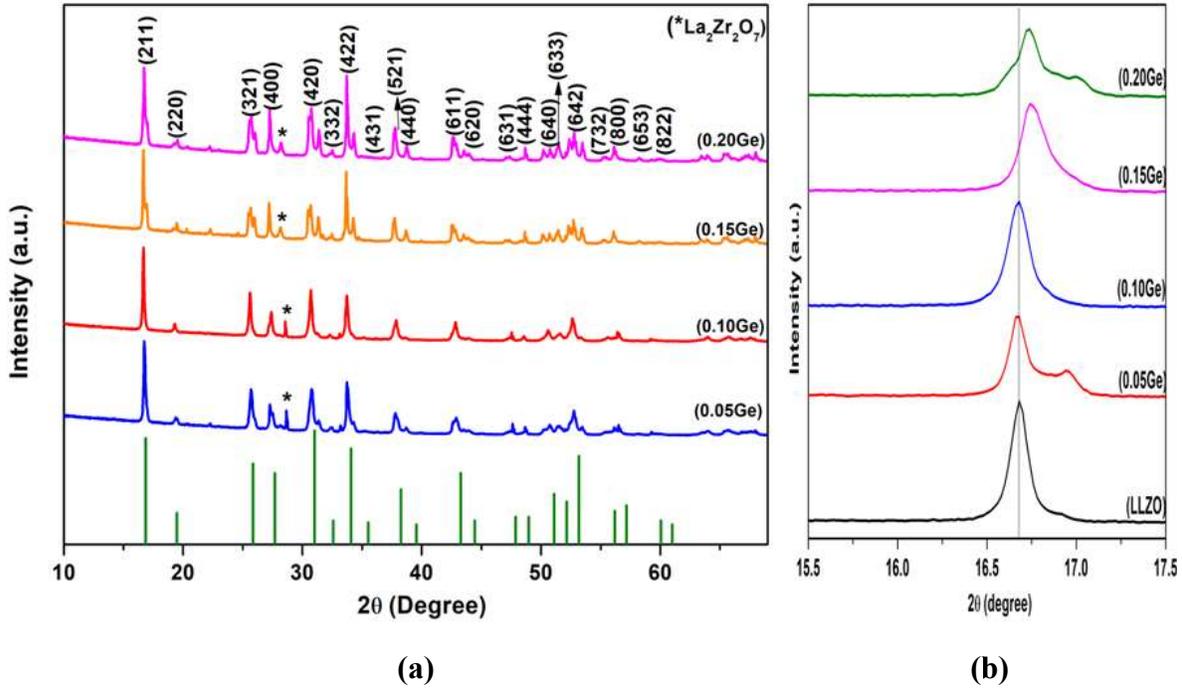

**Fig.1 (a)** X-ray diffraction patterns of $Li_{7-4x}Ge_xLa_3Zr_2O_{12}$ (x=0.05-0.20) and
**b)** Shifting in (211) peak of x= 0.10 Ge

### 3.2 Density Measurement

The density measurement of all the samples of $Li_{7-4x}Ge_xLa_3Zr_2O_{12}$ with x= 0.05-0.20 Ge has been carried out, and the values of density and relative density, along with lattice parameters for all samples, are mentioned in Table 1. From the table, it can be observed that the maximum density and lattice parameter are obtained for the sample with x= 0.10 Ge. Further increase in Ge content leads to decreased density and lattice parameters. This may be due to the substitution of Ge, which has a smaller ionic radius (0.54 $A^0$), at the lattice site of Li, which has a larger ionic radius (0.76 $A^0$). This result agrees with the shifting of the (211) peak due to the shrinking effect, which has been discussed earlier. A similar decrease in lattice constant with density has been reported earlier where $Ge^{4+}$ is substituted in the place of $Zr^{4+}$ [23].

**Table1.** Density and Relative density of Ge substituted $Li_{7-4x}Ge_xLa_3Zr_2O_{12}$.

| x | Density (g/cm³) | Relative density (%) | Lattice constant |
|---|---|---|---|
| 0.05 | 4.23 | 83.10 | 12.9872 |
| 0.10 | 4.4 | 86.44 | 12.9914 |
| 0.15 | 4.11 | 80.74 | 12.9792 |
| 0.20 | 4.09 | 80.35 | 12.9703 |

## 3.3 Morphological and EDS Studies

Fig.2 (a,b,c,d) shows the magnified images of typical surface micrographs of $Li_{7-4x}Ge_xLa_3Zr_2O_{12}$ with Ge varying from 0.05-0.20 mol. In Fig.2 (a), it can be observed that voids are present with irregularities in the size of particles. Fig.2 (b) shows the sample with x=0.10 Ge. It can be observed that, as compared to all other samples, the sample with x= 0.10 Ge has a dense structure with a larger particle size. The grains are well connected with the neighboring grains giving compact nature to the sample. No pores can be seen. This compact arrangement of particles results in a high density of the pellet as compared to other samples. The values of density can be clearly reviewed in Table 2. Fig. (c & d) clearly shows the decrease in density with the increase in Ge content. From the figure, it is clear that, as the amount of Ge increases in $Li_{7-4x}Ge_xLa_3Zr_2O_{12}$, the grain has grown in size, but the void spaces also increase increases the porosity of the sample and results in a lowering of density values. These pores also obstruct the $Li^+$ ion conduction within the material. This may be due to the effect of the viscous flow of sintering that might have occurred in the sample [23]. The elemental mapping of $Li_{6.6}Ge_{0.1}La_3Zr_2O_{12}$ is shown in Fig.3, where grain boundaries can be observed. All the elements viz. La, Zr, and Ge, which are constituents of made samples, are uniformly distributed throughout the sample. Elemental mapping of Ge confirms the insertion of Ge into the lattice. This is also supported by the X-ray diffraction analysis in which shifting of peak occurs due to a change in lattice constant. The EDX spectrum of the sample with 0.10 Ge has been shown in Fig.4 (a). The average particle size was calculated from the histogram mentioned in Fig.4 (b), and it was found to be 3.28 μm.

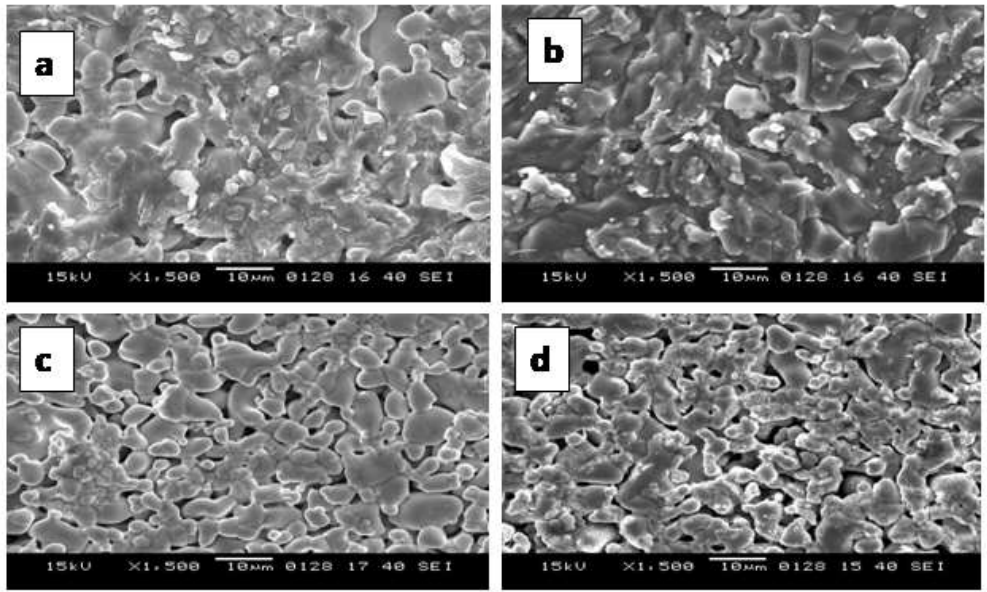

**Fig.2** SEM images of Ge substituted $Li_{7-4x}Ge_xLa_3Zr_2O_{12}$ with x= **a)** 0.05  **b)** 0.10  **c)** 0.15  **d)** 0.20.

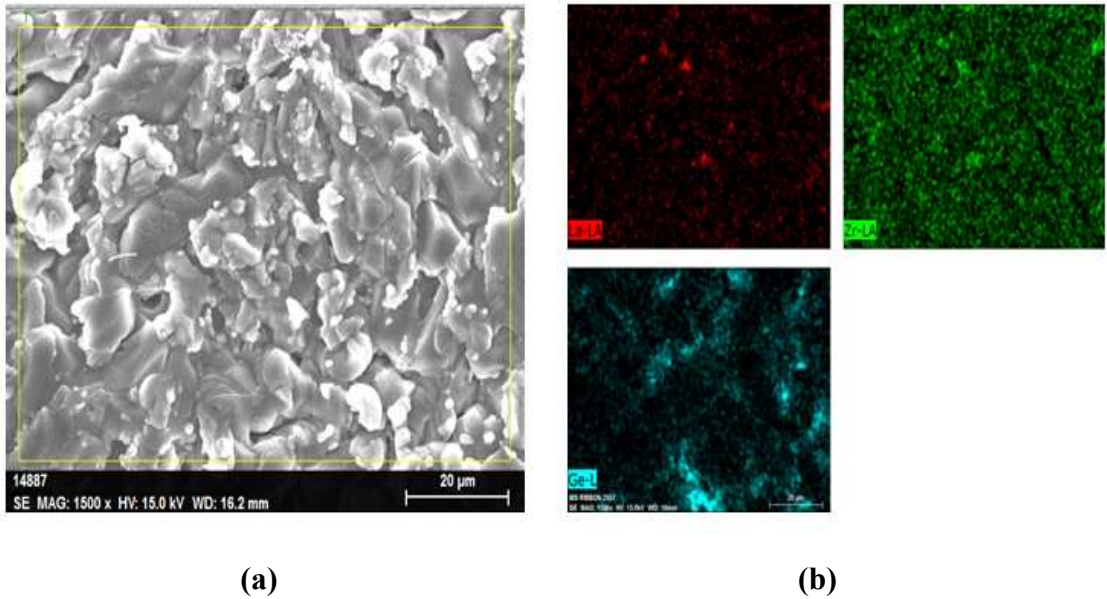

**(a)**                      **(b)**

**Fig.3 (a)** Magnified image of $Li_{6.6}Ge_{0.1}La_3Zr_2O_{12}$    **(b)** Elemental mapping of La, Zr and Ge for $Li_{6.6}Ge_{0.1}La_3Zr_2O_{12}$.

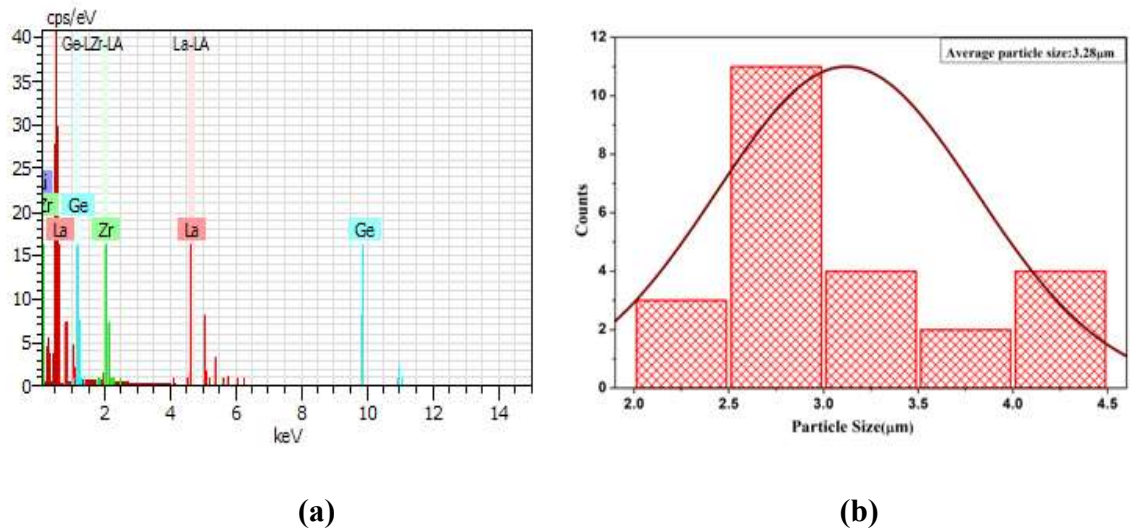

**(a)**                      **(b)**

**Fig.4 (a)** EDX spectrum and **(b)** Particle size distribution of $Li_{6.6}Ge_{0.1}La_3Zr_2O_{12}$.

## 3.4 Conductivity Studies

### 3.4.1. Impedance Plots

The frequency range of 20Hz to 20MHz was used electrical conductivity of $Li_{7-4x}Ge_xLa_3Zr_2O_{12}$ for all the values of x. Fig. 5(a) shows the Nyquist plots for the samples with Ge ranging from 0.05, 0.10-0.20 at $25^0C$. Fig. 5(b) shows the fitted graph for 0.10 Ge substituted LLZO. The ionic conductivity is calculated by using the equation,

$$\sigma = \frac{t}{RA} \tag{1}$$

in which, $\sigma$ is the ionic conductivity, $t$ is the thickness of the sample, $A$ is the area of the electrode, and $R$ is the resistance offered by the sample. The resistance can be calculated from the intercept made by the semicircle on the real axis of Zs' in the high-frequency region. The appearance of a tail in a low-frequency region is because of the Ag electrode's blocking nature. It can be observed from Fig. 5 that the sample with x = 0.10 Ge has a minimum intercept on the real axis offering minimum resistance as compared to other samples. This indicates that substituting $Ge^{4+}$ at $Li^+$ lowers the resistance of LLZO. The sample with 0.10 Ge has maximum conductivity of 7.23 x $10^{-6}$S/cm at room temperature, which is one order higher in magnitude than pure LLZO. This can be explained on the basis of uniformity and increment in grain size, good contact with neighboring grains with no visible pores, high density and stabilized cubic phase. With further increase in Ge content, intercept on real axis shifts towards lower frequency region, giving high resistance and thus decrease in conductivity. This can attributed to formation of a non-conductive $La_2Zr_2O_7$ phase. Thus the high ionic conductivity of 7.23 x $10^{-6}$S/cm at $25^0C$ for 0.10 Ge substituted LLZO makes this ceramic a prominent candidate for battery applications as a solid electrolyte.

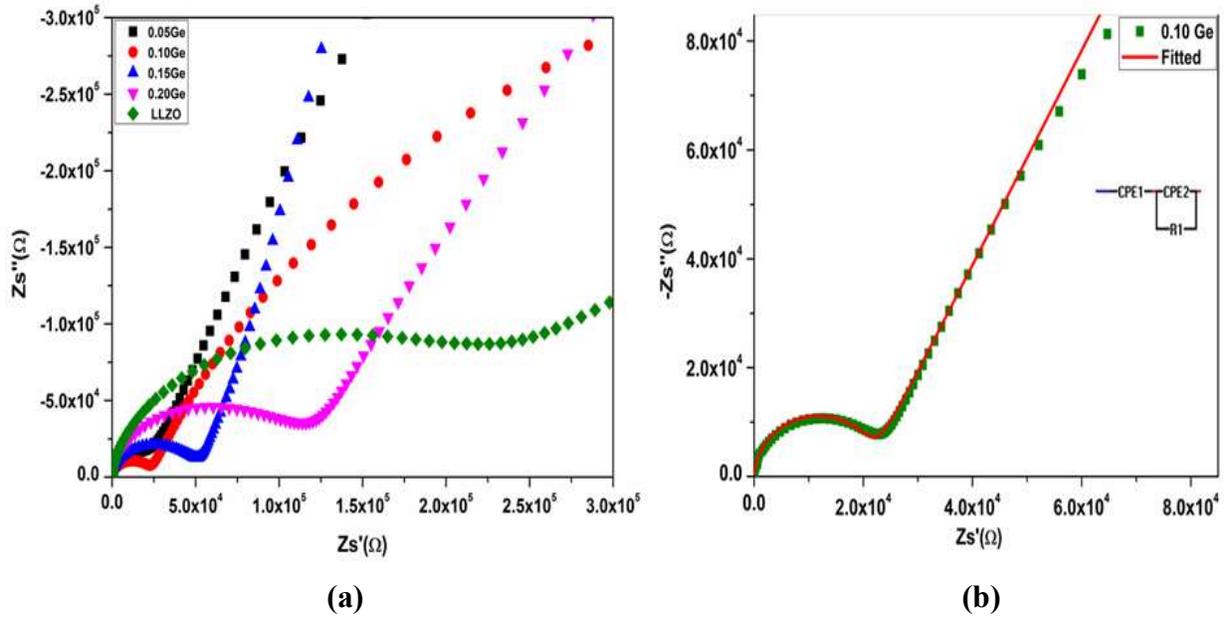

(a)        (b)

**Fig.5 (a)** Impedance plots of $Li_{7-4x}Ge_xLa_3Zr_2O_{12}$ (x=0-0.20) **(b)** Fitted curve for 0.10 Ge.

## 3.4.2. Arrhenius Plots

The Arrhenius plots for $Li_{7-4x}Ge_xLa_3Zr_2O_{12}$ with Ge varying from 0, 0.05– 0.20 mol are shown in Fig.6 (a). This data was collected in the temperature range from $50^0C$ – $150^0C$. The graph plotted using acquired data follows Arrhenius behavior. The Arrhenius equation is given as

$$\sigma(T) = \sigma_0 \exp\left(-\frac{E_a}{K_B T}\right) \quad (2)$$

Where, $E_a$ is the activation energy, $\sigma$ is the conductivity, $\sigma_0$ is the pre-exponential factor, $K_B$ is the Boltzmann constant, and $T$ is the temperature in Kelvin. The activation energy was calculated using this equation for lithium ion conductivity. From the figure, it can be observed that the minimum activation energy ($E_a$) is obtained for x= 0.10 Ge, which also has maximum ionic conductivity at room temperature. The value of activation energy is found to be 0.56 eV for $Li_{6.6}Ge_{0.1}La_3Zr_2O_{12}$. The activation energy values with its corresponding ionic conductivity for all the samples have been tabulated in Table 2. The minimum activation energy for 0.10 Ge containing sample may assign to high relative density. This makes hopping pathways favorable for lithium-ion conduction. Migration of Li-ion is possible through Li point defect as well as Frankel defect between two tetrahedral within the structure according to density functional theory. The distribution of Li vacancy and interstitial lithium in the structure strongly affects Li conductivity [27]. Based on charge neutrality, $Ge^{4+}$ should substitute $Li^+$ and creates three vacancies. Thus the concentration of Li vacancy increases, which eventually leads to an increase in conductivity and minimizes the activation energy. This result is well supported by the earlier report in which replacement of $Al^{3+}$ with $Li^+$ has been done [28]. Fig.6 (b) shows the variation of activation energy and ionic conductivity at $25^0C$ with increasing Ge content in $Li_{7-4x}Ge_xLa_3Zr_2O_{12}$.

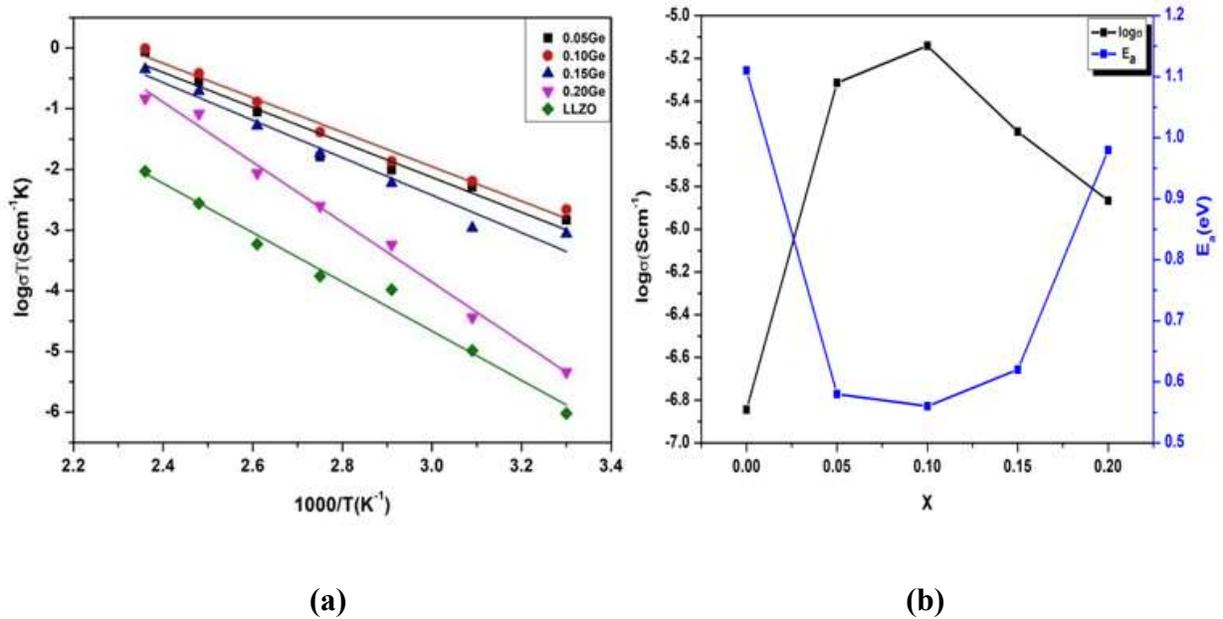

(a)          (b)

**Fig.6 (a)** Ionic conductivity as a function of temperature for x= 0, 0.05, 0.10, 0.15 & 0.20 Ge substituted $Li_{7-4x}Ge_xLa_3Zr_2O_{12}$. **(b)** Variation of conductivity at $25^0C$ and activation energy with varying content of Ge in $Li_{7-4x}Ge_xLa_3Zr_2O_{12}$.

**Table 2.** Ionic Conductivity of Ge substituted $Li_{7-4x}Ge_xLa_3Zr_2O_{12}$.

| x | $\sigma_{bulk}$ (S/cm) at $25^0C$ | $E_a$ (eV) |
|---|---|---|
| 0 | $1.43 \times 10^{-7}$ | 1.11 |
| 0.05 | $4.85 \times 10^{-6}$ | 0.58 |
| **0.10** | **$7.23 \times 10^{-6}$** | **0.56** |
| 0.15 | $2.86 \times 10^{-6}$ | 0.61 |
| 0.20 | $1.36 \times 10^{-6}$ | 0.98 |

### 3.4.3. DC Conductivity

DC conductivity measurement plays an important role in ensuring the presence of electronic conduction in the sample from the calculation of ionic transport number [2]. Fig.7 shows the DC conductivity plots of $Li_{7-4x}Ge_xLa_3Zr_2O_{12}$ with x= 0.05 and 0.10 Ge samples. The whole data was analyzed using a Keithley electrometer up to 300 minutes. For the calculation of ionic transport number, the equation

$$t_i = (\sigma_{total} - \sigma_e)/\sigma_{total} \quad (3)$$

is used where

$$\sigma_{total} = \sigma_{ions} + \sigma_{electrons} \quad (4)$$

It was found to be $> 0.999$ for the specimens with 0.05 and 0.10 Ge. Thus, confirming the predominant ionic conduction within the ceramics.

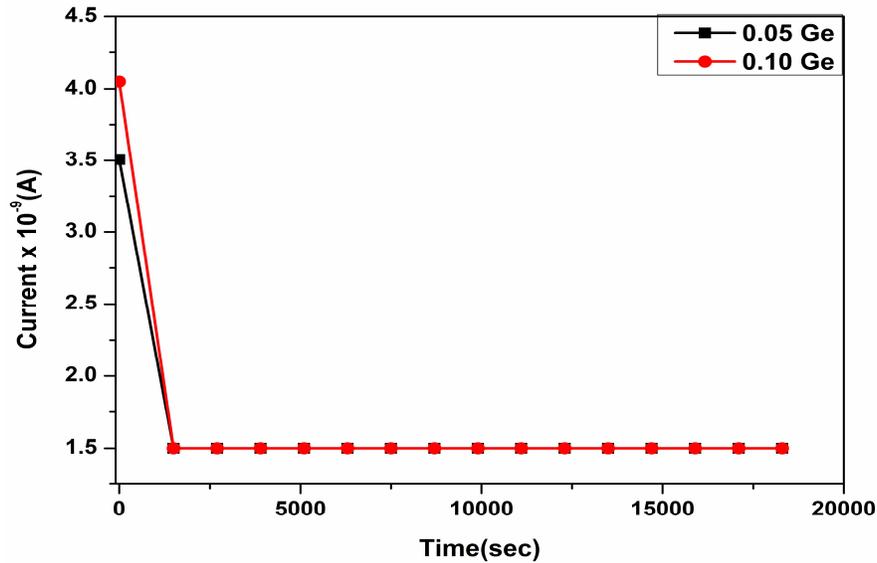

**Fig.7** Variation of DC conductivity with time for transport number measurement.

## 3.5 FTIR Study

FTIR spectroscopy is used to study the presence of functional groups within the material. Various peaks corresponding to the respective groups can be observed in the spectra. Fig.8 depicts the FTIR spectra of all Ge-substituted LLZO samples. The peaks at 563 cm$^{-1}$ and 866 cm$^{-1}$ could be related to Zr-O, whereas the peak observed at 681 cm$^{-1}$ corresponds to the La-O group. Peaks at 1438 and 1504 cm$^{-1}$ are the result of the asymmetric stretching vibration mode of $v_{as}$ (C=O), gives the evidence of the presence of Li$_2$CO$_3$ in the samples[29], [30]. The effect of moisture absorbed from the surrounding by the specimens can be observed from the broad peaks in the region of 3400-3600 cm$^{-1}$ related to OH groups [29].

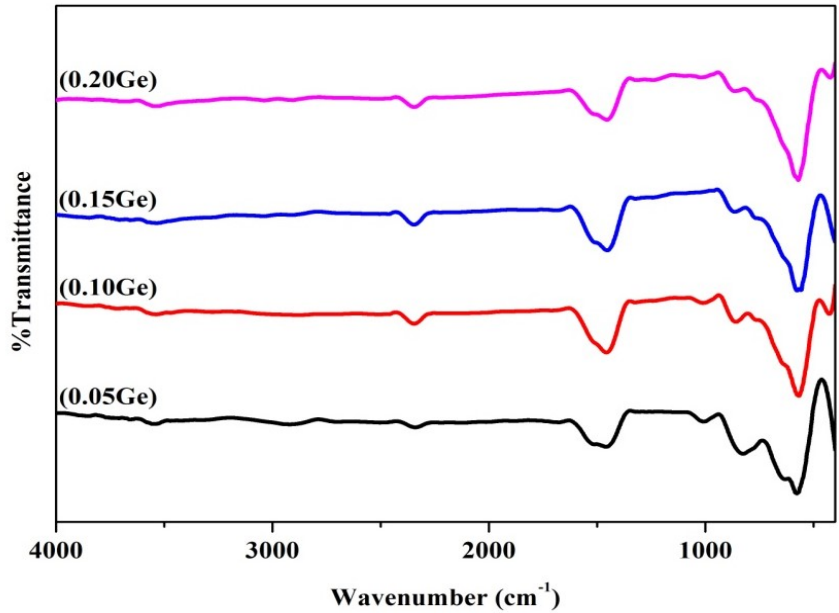

**Fig.8** FTIR spectra of Ge substituted LLZO with x ranging from 0.05 – 0.20.

## 3.6 AC conductivity and modulus study

Two mechanisms are studied to interpret a material's electrical data: ac conductivity and electric modulus [31]. Here in this study, both mechanisms are discussed. Fig. 9 (a) shows the variation of ac conductivity of 0.10 Ge substituted LLZO with frequency. The nature of graph comprises three distinct regions. These are (1) low-frequency region due to polarization at the electrode-electrolyte interface, (2) mid plateau region, which is ascribed to conductivity which is frequency independent and (3) high-frequency region, which shows an increase in conductivity with the frequency. In this high-frequency region, under the influence of an alternating electric field, the movement of ions in the structure is given by Jonscher's universal power law. According to this law, the real part of ac conductivity can be expressed as

$$\sigma(\omega) = \sigma_{dc} + A\omega^n \tag{5}$$

Where, $\sigma(\omega)$ is the total conductivity, $\sigma_{dc}$ is dc conductivity, $A\omega^n$ is the dispersive component of ac conductivity in which $A$ is constant, and $n$ is frequency exponent value. The value of n is physically acceptable in the range of $0 \leq n \leq 1$ and subjected to the interaction of the ions. In Table 3, the values of $n$ are listed for 0.10 Ge substituted LLZO for different temperatures. From that, it can be observed that with the increase in temperature, the value of n increases which can be well explained based on a theoretical model named quantum mechanical tunneling (QMT). Also, the increase in conductivity with the temperature occurs due to the mobility of ions. Temperature-dependent frequency exponent obtained in the quantum mechanical tunneling model framework assumes that charge carriers form non-overlapping small polaron. In this model, polaron hopping energy and characteristic relaxation time are used to calculate it [32].

Fig.9 (b) shows the scaled temperature-dependent spectrum of the real part of ac conductivity of 0.10 Ge. It can be observed from the graph that the conductivity spectra at different temperatures almost merge into a single master curve which is an important feature of temperature independent relaxation process under conductivity formalism. At higher frequencies, deviation in the curve can be observed which is possibly due to structural peculiarities in the specimen caused by different conduction pathways. In comparison, the variation in the superimposition of conductivity spectra at lower frequencies gives compositional dependence of the material. It may be ascribed to the polarization effect due to the electrode-electrolyte interface [31], [32].

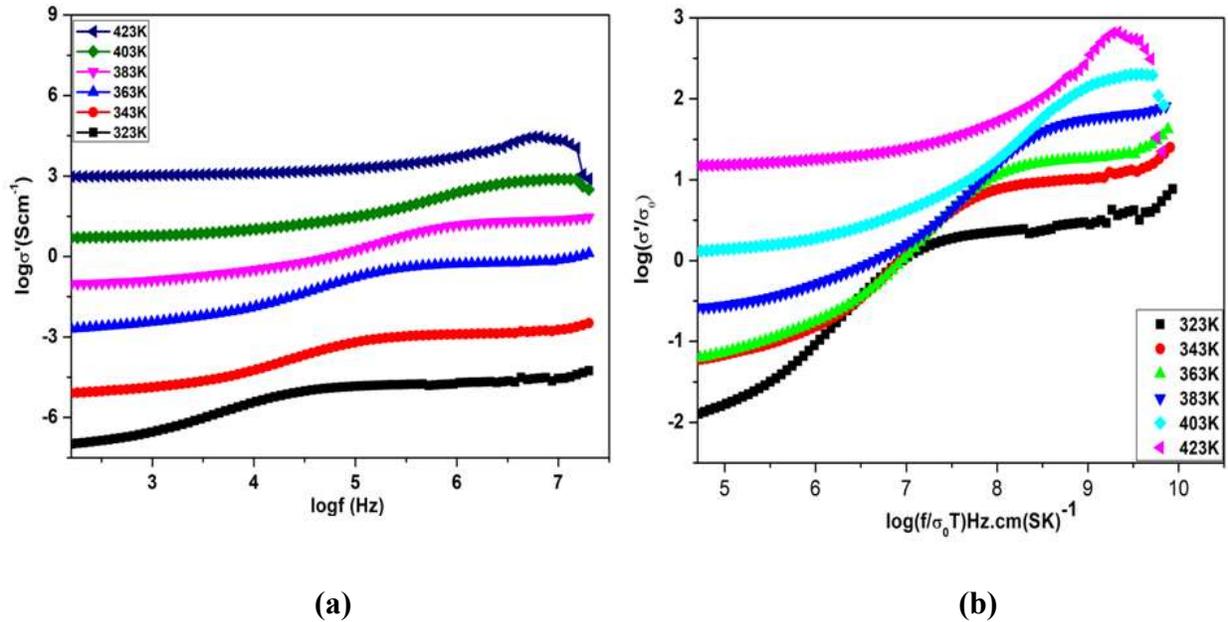

(a)          (b)

**Fig.9 (a)** Frequency-dependent ac conductivity of 0.10Ge substituted $Li_{7-4x}Ge_xLa_3Zr_2O_{12}$.
**(b)** The scaling behaviour of 0.10 Ge substituted $Li_{7-4x}Ge_xLa_3Zr_2O_{12}$ at different temperatures.

The electric modulus study is used to build the relation between conductivity and relaxation of mobile ions in conducting solids. According to this, complex electric modulus ($M^*$) is the reciprocal of complex permittivity ($\varepsilon^*$), and the equation is given by

$$M^* = \frac{1}{\epsilon^*} = \frac{(\epsilon' - j\epsilon'')}{|\epsilon^*|^2} = M' + jM'' \tag{6}$$

Where, $M'$, $M''$, $\epsilon'$, $\epsilon''$ are real and imaginary parts of complex modulus and complex permittivity respectively [31], [33]. Fig. 10 (a) shows variation of $M''$, with the frequency for 0.10 Ge substituted LLZO at different temperatures. This represents energy loss under an applied electric field. The peak in the spectrum also represents the conductivity relaxation[34]. From the figure, two regions can be identified, i.e. the region of frequency below $M''$ peak and frequency region above $M''$ peak. These regions show the range where charge carriers are mobile on long and short distances. For further evaluation, relaxation time is calculated using frequency corresponding to $M_{max}''$. The plots of $M''/M_{max}$ Vs $f/f_{max}$ known as modulus scaling, are shown in Fig. 10 (b). Here the overlapping of curves can be observed for the distinct temperature range, which indicates temperature-independent conduction. Fig. 10 (c) represents the variation of $M''/M_{max}$ Vs $f/f_{max}$ of all the Ge substituted LLZO. In this figure, the non-overlapping curves indicate that the conduction phenomenon depends on the composition. This may attribute to the insertion of Ge within the sample. Variations in relaxation time with temperatures can be seen in Fig. 10 (d). The nature of the graph follows Arrhenius law, i.e.

$$\tau = \tau_0 \exp\left(\frac{E_a}{K_B T}\right) \tag{7}$$

where, $\tau_0$ is a pre-exponential factor, $E_a$ is the activation energy, $K_B$ is Boltzmann constant, and T is the temperature in Kelvin. The values of activation energies for $E_a(\tau)$ and $E_a(\sigma)$ are found to be 0.52 eV and 0.56 eV, respectively. There is a small change in both energy values. This small difference does not affect any properties and suggests that the same type of charge carrier is responsible for both the processes of conduction and relaxation.

**Table3**. Frequency exponent values of 0.10 Ge substituted $Li_{7-4x}Ge_xLa_3Zr_2O_{12}$.

| Temperature ($^0$C) | Frequency exponent (n) |
|---|---|
| 150 | 0.62 |
| 130 | 0.59 |
| 110 | 0.57 |
| 90 | 0.52 |
| 70 | 0.50 |
| 50 | 0.47 |

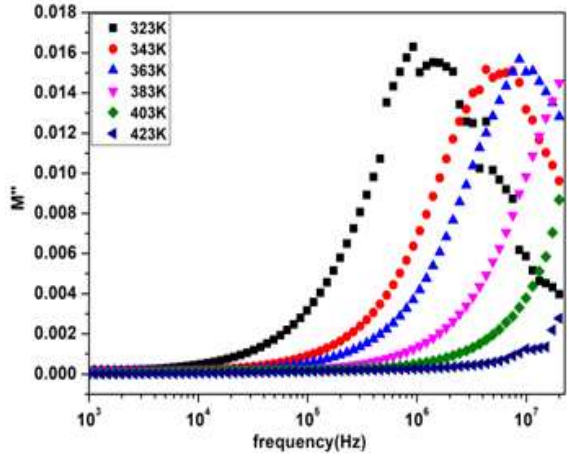
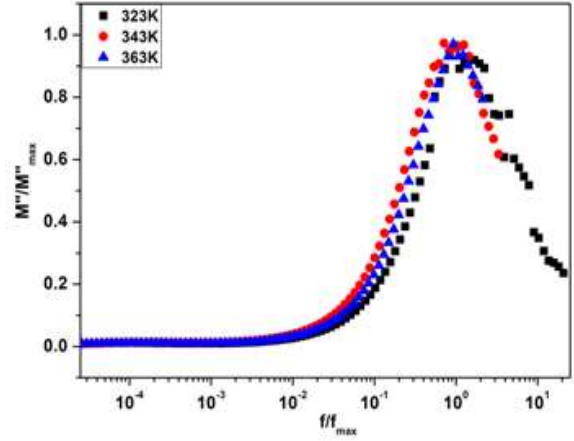

(a)

(b)

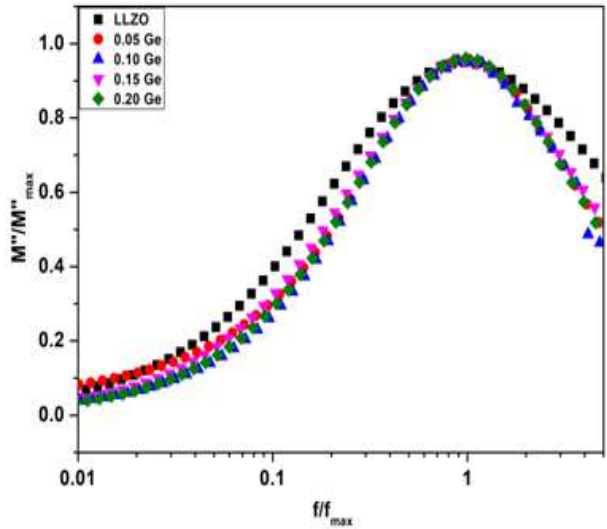
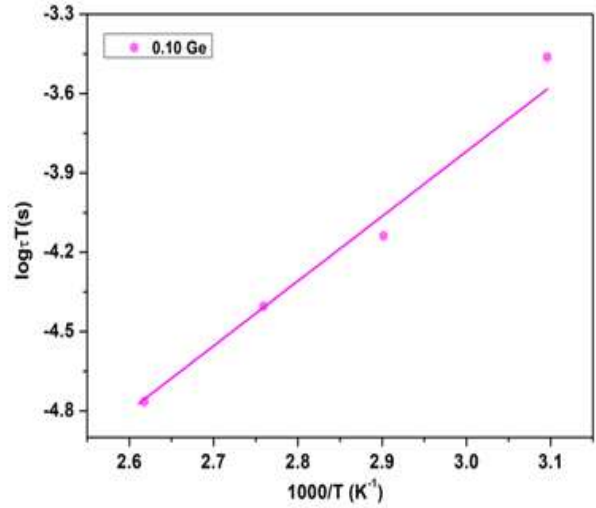

(c)

(d)

**Fig.10 (a)** Imaginary part of electric modulus as a function of frequency and temperature for 0.10 Ge substituted $Li_{7-4x}Ge_xLa_3Zr_2O_{12}$, **(b)** Normalized modulus plots ($M''/M''_{max}$) Vs log ($f/f_{max}$) for 0.10 Ge substituted $Li_{7-4x}Ge_xLa_3Zr_2O_{12}$ at different temperatures, **(c)** Normalized modulus plots ($M''/M''_{max}$) Vs log ($f/f_{max}$) for x= 0 - 0.20 Ge substituted $Li_{7-4x}Ge_xLa_3Zr_2O_{12}$ at 25 $^0$C, **(d)** Variation of relaxation time with temperature for 0.10 Ge substituted $Li_{7-4x}Ge_xLa_3Zr_2O_{12}$.

## 4. Conclusions

The garnet type $Li_{7-4x}Ge_xLa_3Zr_2O_{12}$ (x= 0.05-0.20) has been prepared by the conventional solid-state reaction method. The cubic phase has been confirmed by XRD characterization. The density measurement and SEM analysis confirmed the compactness and uniformity in grain size for x= 0.10 Ge substituted $Li_{7-4x}Ge_xLa_3Zr_2O_{12}$. The elemental mapping for this sample reveals the uniform distribution of Ge in the specimen with an average particle size of 3.28 μm. There is an enhancement in an ionic conductivity by one order of magnitude for 0.10 Ge at room temperature with minimum activation energy. This has been attributed to the substitution of supervalent dopant for lithium, which increases the pathways for lithium-ion conduction. The FTIR study confirmed the existence of the functional groups in the samples. The ac conductivity and modulus study confirmed that, both the conduction and relaxation processes are due to the same kind of charge carrier and the relaxation process is temperature independent. Thus, 0.10 Ge substituted $Li_{7-4x}Ge_xLa_3Zr_2O_{12}$ having ionic conductivity of 7.23 x $10^{-6}$ S/cm at room temperature and activation energy of 0.56 eV, is a prominent candidate for the solid electrolyte.

## Acknowledgments

One of the authors would like to express sincere appreciation to VNIT, Nagpur, for providing a Ph.D. fellowship. The author appreciate the support of DST FIST project number SR/FST/PSI/2017/5(C) for the XRD facility provided by the Department of Physics at VNIT, Nagpur.

**Statements & Declarations**

**Author Contributions**

MA: Material preparation, data analysis, writing original draft, conceptualization, editing, proof reading.

AD: Editing, supervising, conceptualization.

**Funding**

The authors declare that no funds, grants, or other support were received during the preparation of this manuscript.

**Competing Interests**

The authors have no relevant financial or non-financial interests to disclose.